\documentclass[twocolumn,showpacs,preprintnumbers,amsmath,amssymb,prb,a4paper,english]{revtex4}

\usepackage{amsmath}
\usepackage{amsfonts}
\usepackage{amssymb}
\usepackage{lipsum,babel}
\usepackage{array}
\usepackage{adjustbox}
\usepackage{multirow}
\usepackage{epstopdf}

\newcommand{\ersample}{$^{167}\text{Er}^{3+}\text{:Y}_2\text{SiO}_5$}
\newcommand{\er}{$^{167}\text{Er}^{3+}$}
\newcommand{\YSO}{$\text{Y}_2\text{SiO}_5$}

\begin{document}
\title{Hyperfine interactions of $\text{Er}^{3+}$ ions in \YSO: \\ electron paramagnetic resonance in a tunable microwave cavity}

\author{Yu-Hui Chen}
\author{Xavier Fernandez-Gonzalvo}
\author{Sebastian P. Horvath}
\author{Jelena V. Rakonjac}
\author{Jevon J. Longdell}

\affiliation{The Dodd-Walls Centre for Photonic and Quantum Technologies, Department of Physics, University of Otago, 730 Cumberland Street, Dunedin, New Zealand.}

\date{\today}

\begin{abstract}
The hyperfine structure of the ground state of erbium doped yttrium orthosilicate is analyzed with the use of electron paramagnetic resonance experiments in a tunable microwave resonator. This work was prompted by the disagreement between a recent measurement made in zero magnetic field and a previously published spin Hamiltonian which. 
 The ability to vary magnetic field strength, resonator frequency, and the orientation of our sample enabled us to monitor how the frequencies of hyperfine transitions change as a function of a vector magnetic field. We arrived at a different set of spin Hamiltonian parameters, which are also broadly consistent with the existing data. We discuss the reliability of our new spin Hamiltonian parameters to make predictions outside the magnetic field and frequency regimes of our data. We also discuss why it proved to be difficult to determine spin Hamiltonian parameters for this material, and present data collection strategies that improve the model reliability.
\end{abstract}

\maketitle
 
\section{Introduction}

Due to long optical and hyperfine coherence times, rare earth doped crystals are considered to be a very promising candidates in the quest for applications in quantum memories and quantum information. These materials have been shown to have optical coherence times on the order of milliseconds and allow for tailoring of the inhomogeneous linewidth using external electric or magnetic fields \cite{Thiel2011}.  By transferring the excitations to hyperfine spin energy levels, it has been demonstrated that the coherence time of rare-earth doped crystals can be extended to six hours \cite{Zhong2015}. The physics behind this extension is to find a specific magnetic field so that the hyperfine transitions have zero first-order Zeeman (ZEFOZ) shift and are only sensitive to second order magnetic field fluctuations \cite{Fraval2005, Zhong2015}. To utilize this technique, one requires an accurate model of the hyperfine structure for the rare-earth dopant. 

Among the rare earth ions, erbium (Er) has a unique optical transition located at 1.5 $\mu$m that makes it compatible with optical fibres, and Er doped crystals have both optical life times and coherence times of the order of milliseconds \cite{Thiel2011,Macfarlane1997, Bottger2006}. For this reason,
Er based optical quantum memories are strong candidates for inclusion in a future optical fiber based quantum network. Moreover, erbium has a stable isotope $^{167}\text{Er}$ (natural abundance 22.94\%) with a nuclear spin of $I=7/2$, which results in a hyperfine structure extending over a 5\,GHz range at zero magnetic field. This has prospects for achieving ultra-long coherence times \cite{McAuslan2012}, as well as enabling the development of quantum microwave-to-optical converters \cite{Williamson2014a, Chen2016}.

For this reason, there has been considerable effort devoted to understanding the hyperfine structure of $^{167}$Er doped yttrium orthosilicate (\ersample). This includes the identification of effective three-level $\Lambda$ systems \cite{Baldit2010}, as well as, more recently, the observation of a ground-state coherent Raman process with a coherence time of 50 $\mu$s \cite{Hashimoto2016}, which was limited by electron spin-spin interactions.  This limitation has been circumvented by experiments employing strong external magnetic fields, yielding a coherence time of 1.3 seconds for hyperfine transitions in \ersample \cite{Rancic2016}, and thus demonstrating the practical viability of quantum memories at 1.5\,$\mu$m. In order to guide such developments using accurate theoretical models, in addition to enabling the above outlined ZEFOZ technique, it is crucial to have an accurate understanding of the hyperfine structure of Er doped yttrium orthosilicate \cite{Asatryan2002, Jerzak2003, Guillot-Noel2006, Sun2008, Marino2016}.

The hyperfine struncture of the ground state $^4I_{15/2}Z_1$ of Er doped \YSO \ was first measured by standard electron paramagnetic resonance (EPR) experiments at 9.5\,GHz by Guillot-N\"oel \textit{et al.} \cite{Guillot-Noel2006}. By analyzing the angular variations of the eight allowed and some forbidden hyperfine transitions the spin Hamiltonian parameters were determined. However, the predicted ground-state energy levels of \ersample\, from this original set of parameters resulted in some discrepancies with zero-field measurements \cite{Chen2016}. 

In this paper, the ground-state spin Hamiltonian parameters of \ersample \ are determined by EPR experiments with a tunable loop-gap resonator.  In contrast to previous EPR measurements using a fixed-frequency microwave resonator \cite{Guillot-Noel2006}, both the resonator frequency as well as the applied magnetic field were varied to yield two dimensional EPR scans, allowing one to monitor how the frequency of a particular hyperfine transition varies with an applied magnetic field. Because of the large number of parameters, simulated annealing \cite{Aster2011} was used to find a set of parameters that best fit the observed spectra. Published zero-field EPR data \cite{Chen2016} was then introduced into the calculation to refine the fitted Hamiltonian. In order to determine the uncertainties in the spin Hamiltonian parameters, the Markov-Chain-Monte-Carlo technique \cite{Aster2011} was used to sample the posterior probability distribution.
  
\section{Erbium doped \YSO}

When using rare earth ion doped crystals for quantum information applications, ideally the host crystal would be free of nuclear and electron spins so that there is no magnetic pertabations in the local fields for the doped ions. While no such host crystal has been demonstrated, \YSO\ has low nuclear-spin fluctuations. Yttrium (100\% $^{89}$Y) has a nuclear magnetic moment of only -0.137$\mu_n$, Si has one magnetic isotope of abundance of 4.6\% with a moment of -0.554$\mu_n$, and O has one magnetic isotope of abundance of 0.04\% with -1.89$\mu_n$ \cite{Baldit2010}, where $\mu_n$ is the nuclear magneton. Very long nuclear spin coherence times have been demonstrated for both praseodymium \cite{Fraval2005} and europium \cite{Zhong2015} dopants in \YSO\, and on the basis of previously published spin Hamiltonian parameters \cite{Guillot-Noel2006} long spin coherence times can also be expected for \ersample\ \cite{McAuslan2012}.  Attempts to utilize \ersample\ in quantum information have been hampered by the low symmetry of the rare earth dopant, although the low symmetry does provide benefits for building a $\Lambda$ system. \YSO\ has a monoclinic structure with C$^6_{2h}$ space group and two crystallographic sites of $C_1$ symmetry, site 1 and site 2, where Er$^{3+}$ ions can substitute for yttrium ions. Here we follow the definition of site 1 and site 2 where $^4\textit{I}_{15/2}Z_1 \leftrightarrow {^4\textit{I}_{13/2}Y_1} $ transition is at 1536 nm for site 1 and 1539 nm for site 2 as in the references \cite{Macfarlane1997, Guillot-Noel2006, Sun2008}. For each crystallographic site, there are two magnetically inequivalent ion subclasses for a magnetic field in an arbitrary direction, with the exception of when the magnetic field is aligned along the crystals $C_2$ (or ``$\mathbf{b}$'') axis or perpendicular to it \cite{Sun2008}. As \er ions have a nuclear spin of $I$\,=\,7/2 and an effective electronic spin of $S$\,=\,1/2, there are 16 hyperfine energy levels for the ground state, even in the absence of an external magnetic field. These hyperfine splittings can be represent by the following spin Hamiltonian \cite{A.Abragam1951}
\begin{equation}
H=\mu_e \mathbf{B} \cdot \mathbf{g} \cdot \mathbf{S} + \mathbf{I} \cdot \mathbf{A} \cdot \mathbf{S} +\mathbf{I} \cdot \mathbf{Q} \cdot \mathbf{I} -\mu_n g_n \mathbf{B} \cdot \mathbf{I} ,\label{eq:spinHam}
\end{equation} 
where $\mu_e$ is the Bohr magneton, $\mathbf{B}$ the applied magnetic field, $\mathbf{g}$ the Zeeman $g$-matrix, $\mathbf{A}$ the hyperfine matrix, $\mathbf{Q}$ the electric quadrupole matrix, $\mu_n$ the nuclear magneton, and $g_n=-0.1618$ is the nuclear $g$ factor. However, due to the low point-group symmetry of the \YSO \ crystal, the $\mathbf{g}$, $\mathbf{A}$, and $\mathbf{Q}$ matrices have noncoincident principal axes. Therefore, not only their principal values, but also their individual corresponding Euler angles need to be determined, which makes finding the spin Hamiltonian parameters difficult. 

\section{Experimental Setup}

Our sample is a cylindrical \ersample \  crystal supplied by Scientific Material Inc, which has a length of 12\,mm and a diameter of 4.95\,mm. In this sample, 50 parts per million of the Y$^{3+}$ ions are substituted by isotopically pure \er\ ions.
The coordinate system we will use in this paper uses the crystallographic $\mathbf{b}$ axis, in addition to the two other optical extinction axes $\mathbf{D_1}$, $\mathbf{D_2}$, which are orthogonal \cite{Sun2008}. Orientation of the sample was carried out by Scientific materials with the crystal cut so that the $\mathbf{b}$ axis was  along the longitudinal axis of the cylinder, and the $\mathbf{D_2}$ axis was identified by a flat cut of the curved surface.

\begin{figure}
\hspace*{-0.5cm}  
\includegraphics[width=0.45\textwidth]{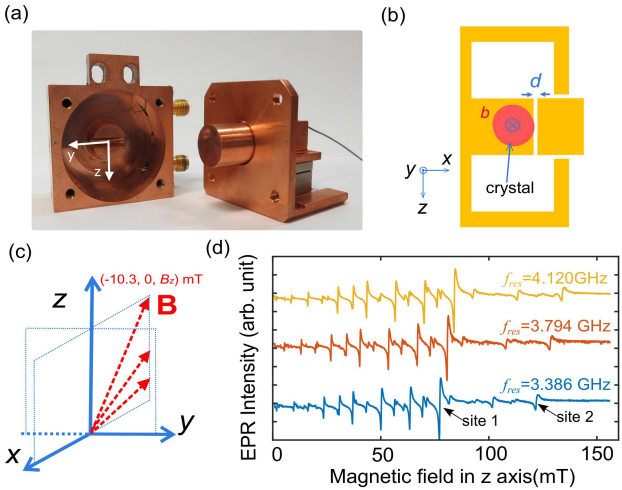}
\caption{ Experimental setup. (a) Photo of our tunable loop-gap resonator. Left, body of our cavity; right, a cap and a plunger. $y$ and $z$ indicate the corresponding directions of our magnetic field. (b) Schematic picture of the tunable resonator. The resonant frequency can be tuned by moving the plunger to change the gap, $d$. Coordinate systems of our crystal and our magnetic field are also shown. (c) Schematic of the applied magnetic fields. Shown here is the scanning of $B_z$ for $B_x=10.3$\,mT and $B_y=0$\, mT. (d) Typical EPR lines at three different resonator frequencies. The lines are vertically shifted for clarity. 
\label{fig:setup} }
\end{figure}

The loop-gap resonator, which is made from oxygen free cooper, is shown in FIG.~\ref{fig:setup}. The \ersample \ sample sits in the central hole of the resonator. The tunable resonator is based on a three-loop two-gap configuration, where the resonant frequency can be finely tuned by changing the width of the gap $d$, as illustrated in FIG.~\ref{fig:setup}(b) and further details can be found in reference \cite{Chen2016}. Our cavity and sample assembly were cooled to 4.3\,K using a homebuilt cryostat (cooling head: Cryomech PT405). At this temperature, the resonator typically showed Q factors of $6\times 10^3$ and was tunable from 3\,GHz to 5.5\,GHz.  

The magnetic field was supplied by a custom high temperature superconductor vector magnet from HTS-110 Ltd. It could provide up to $\pm312$\,mT in one direction, $z$, and $\pm10.3$\,mT in the other two, $x$ and $y$. The $y$ axis of the magnet is aligned along the $\mathbf{b}$ axis of our sample, allowing the $\mathbf{D_1}$ and $\mathbf{D_2}$ axes to be placed anywhere in the $x-z$ plane by rotating the sample inside the cylindrical hole of the microwave resonator during the sample mounting process. Each orientation of the crystal can be described by the angle $\theta$ between $\mathbf{D_1}$ and $z$, which is defined as positive when the rotation from $\mathbf{D_1}$ to $z$ along the $\mathbf{b}$ axis is anti-clockwise. Note that our measurements of the magnetic field needed to make the two magnetically inequivalent subclasses degenerate showed a $1.0^\circ$ misalignment between the $x-z$ plane and the $\mathbf{D_1-D_2}$ plane, which were taken into account during the experiment and the fitting process. The values for $\theta$ were first measured with the help of a camera to an accuracy of $\pm 3^\circ$, which were allowed to vary during the fitting process; the resulting values for $\theta$ were in a range of values consistent with the photographic measurements. 

EPR spectra were taken with fixed $x$ and $y$ fields and sweeping the current applied to the $z$ coils. For each orientation ($\theta$) of the sample, and each cavity frequency, we swept $B_z$ from $0$ to $300$\,mT with $(B_x, B_y)$ either $(0,\delta(B_z))\,$mT or one of $(\pm10.3, \pm10.3)$\,mT, where $\delta(B_z)$ was introduced to address the small misalignment between the $x-z$ plane and the $\mathbf{D_1-D_2}$ plane. The applied magnetic fields are illustrated in FIG.~\ref{fig:setup}(c)

\begin{figure*}
\includegraphics[width=0.85\textwidth]{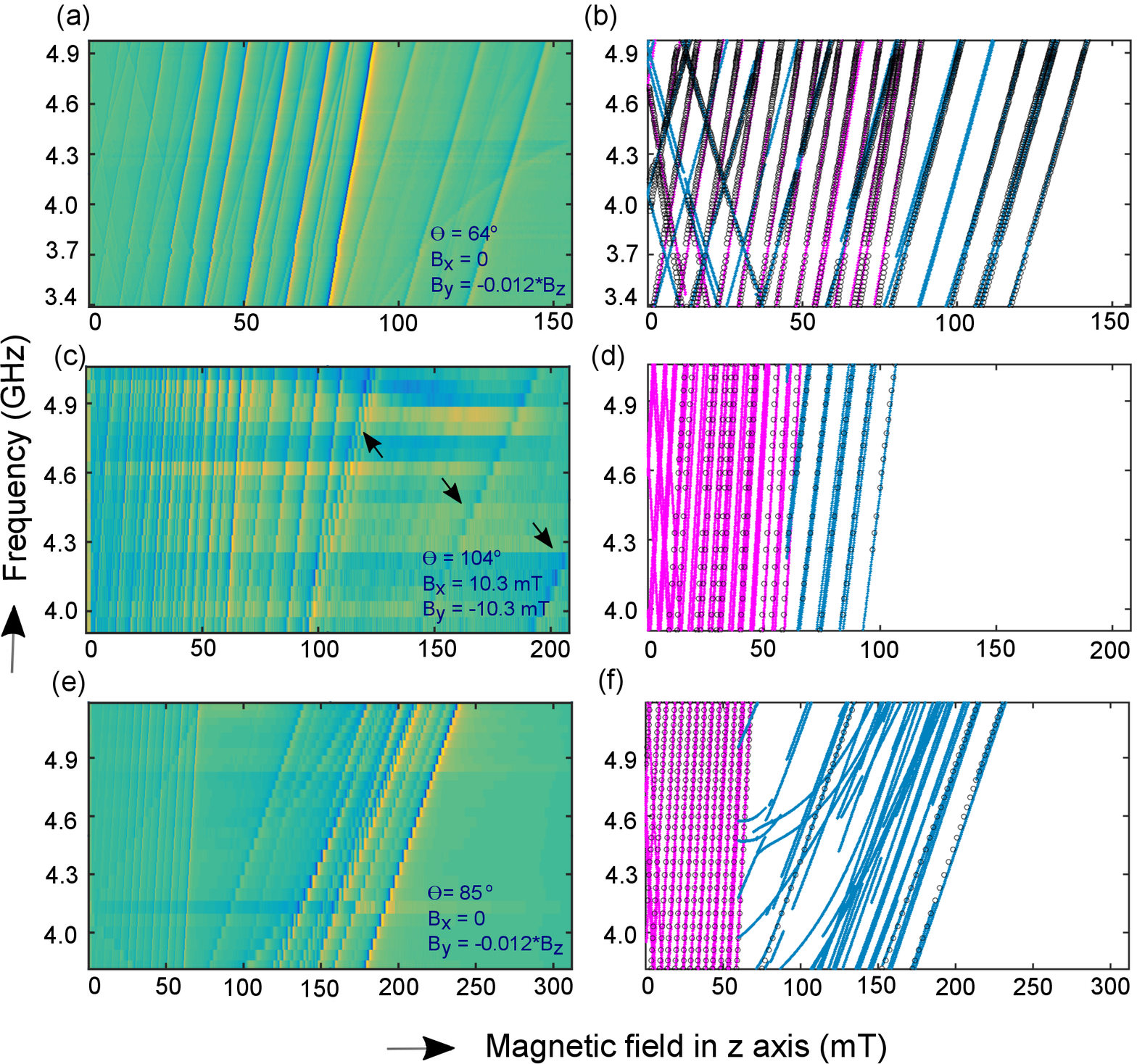}
\caption{ Experimental EPR data and calculated transition lines. In the color plots, each horizontal slice is a normalized EPR spectrum as a function of $B_z$, while the applied $B_x$ and $B_y$ are indicated by the corresponding insets. (a) is the measured EPR spectra at $\theta = 64.04^\circ$. The scanned resonator frequency range is 3388.6\, MHz to 4975.6\,MHz. (c) is the measured EPR spectra at $\theta = 104.04^\circ$. The scanned resonator frequency range is  3960.6\, MHz to 5058.8\,MHz. (e) is the measured EPR spectra at $\theta = 84.96^\circ$. The scanned resonator frequency range is  3811.3\, MHz to 5184.8\,MHz. The circle dots in (b), (d), and (f) are the experimental EPR peaks extracted from (a), (c), and (e), respectively. On top of the experimental points, the pruple lines are the EPR lines for site 1 from calculations using the parameters in Table \ref{tab:vals} and blue lines are for site 2. \label{fig:spectra}}
\end{figure*}

The EPR transitions were detected using a frequency modulation approach that we have reported earlier \cite{Fernandez-Gonzalvo2015a}. Shown in FIG.~\ref{fig:setup}(d) are EPR lines for three different resonator frequencies. By moving the plunger, we were able to gradually tune the frequency of our resonator. We therefore obtained five two-dimensional (2D) EPR scans for each $\theta$ and were able to monitor how the frequencies of particular hyperfine transitions changed as a function of magnetic field.

\section{Results}
Figure \ref{fig:spectra} shows some representative EPR spectra from our experiments. The readers are referred to the Supplemental Material\cite{SupplMat} for a complete set of data. As the absolute strengths of the EPR signals differ  widely according to the resonance frequency of the microwave cavity, which is a result of the frequency dependence of the cable loss and the coupling strength of the antenna, every slice of each color figure in FIG.~\ref{fig:spectra} is an EPR spectrum normalized to their individual maxima. Figures \ref{fig:spectra}(a) and \ref{fig:spectra}(b) were the data for $\theta = 64.0^\circ$, (c) and (d) are for $\theta = 104.0^\circ$, and (c) and (d) are for $\theta = 85.0^\circ$. 

In FIG.~\ref{fig:spectra}, sudden color changes from blue to yellow indicate EPR peaks with their dispersive lineshapes. 
Two sets of `parallel' lines with different `slopes' can be identified in FIG.~\ref{fig:spectra}. They correspond to two different effective $g$ factors. Based on the previous knowledge of the $g$ factor from literature \cite{Guillot-Noel2006, Sun2008}, we can easily distinguish which line belongs to site 1 or site 2 and then input them into the fitting procedure, as guided by the purple lines and the blue lines. In general, we observed more than eight `allowed' transitions in both site 1 and site 2, and the strengths of the EPR signals are on the same order of magnitude. This is because the hyperfine states are highly mixed states at the this relatively low magnetic field.

In FIG.~\ref{fig:spectra}(c) and (d), there are several lines which cannot be ascribed to either site 1 or site 2. See, for example, the lines marked by the black arrows in FIG.~\ref{fig:spectra}(c). They also appear to have hyperfine transitions and are anisotropic; we attribute them to impurities in the \YSO. Such impurities have also been previously observed in \YSO \ crystals \cite{Goryachev2014a}.

The fitting of the spin Hamiltonian parameters was based on our 2D EPR scans. In general, the measured EPR spectra of \er \ are composed of EPR signals from two inequivalent magnetic subclasses of both site 1 and site 2, the spin Hamiltonian parameters of which are related by a $180^\circ$ rotation along the $\mathbf{b}$ axis in \YSO \cite{Sun2008}. That is to say, we normally have four sets of hyperfine lines in each of our 2D EPR scans. However, $\theta$ was chosen to give significant different effective $g$ factors to site 1 and site 2, and thus it was easy to categorize the measured EPR lines to site 1 and site 2 prior to our fitting procedure. 

The spin Hamiltonian parameters \textbf{g}, \textbf{A}, and \textbf{Q} were chosen to be symmetric as in previous measurements \cite{Guillot-Noel2006} and parameterised by their principal values and Euler angles to describe their directions. The Euler angles here are chosen to represent an intrinsic rotation sequence of $z-x'-z''$ from the right-handed $(\mathbf{D_1}, \mathbf{D_2}, \mathbf{b})$ system to the right-handed coordinate system of the principal axes. For example, let matrix $\mathbf{M}$ be defined in $(\mathbf{D_1}, \mathbf{D_2}, \mathbf{b})$ and $\mathbf{M}_{p}$ is defined according to its principal axes. Following this definition, we have
\begin{equation}
\mathbf{M}=R^\text{T} \cdot \mathbf{M}_{p} \cdot R
\end{equation}  
where the rotation matrix $R = R_z(\gamma) \cdot R_x(\beta) \cdot R_z (\alpha)$, and $R_z(\gamma), R_x(\beta), R_z (\alpha)$ represent rotations of $\alpha$, $\beta$, and $\gamma$ along $z$, $x'$ (the $x$ after the first elemental rotation), and $z''$ (the $z$ after the second rotation) axes. In our fitting, the axes with the biggest principal values are chosen to be along $z$, and the other principal axes are allowed to vary. For both site 1 and site 2, the principal axes of \textbf{g}, \textbf{A}, and \textbf{Q} are not coincident. So we had to determine six parameters for \textbf{g}, six for \textbf{A}, and five for \textbf{Q} which is a traceless matrix. Together with four $\theta$s (sample orientations), this results in 38 parameters in total to be determined in our fitting.

From our 2D EPR scans, we can extract the frequencies of hyperfine transitions as a function of $B_z$, i.e., points as $(f, \, B_z)$, where the hyperfine transition frequencies and resonator frequencies satisfy $f_{\text{tran}}=f_{\text{res}}=f$. In principal, a misfit function can be defined by summing the variances of either $B_z$ or $f$. Because $f_{\text{res}}$ can be measured very precisely (the precision can be up to kHz, and it is limited by the noise of the microwave detector), it is better to use $f$ as arguments and $B_z$ as dependent variables to define a misfit function. The misfit function is then defined as
\begin{equation}
\text{misfit} = \sum_i^N ({B_{z,i}^{\text{cal}}} - {B_{z,i}^{\text{exp}}} )^2
\label{eqs:misfit}
\end{equation} 
where $B_{z,i}^{\text{cal}}$ are the calculated $B_z$ at one determined peak $f_{i}$, $B_{z,i}^{\text{exp}}$ are the corresponding experimental $B_z$ values, $i$ is the index of our EPR peaks, and $N$ is the total number of EPR peaks that were used in the fitting. It is easy to calculate the hyperfine transition frequencies for a given magnetic field (by diagonalization of the total Hamiltonian), but the inverse problem, getting $B_{z,i}^{\text{cal}}$ for a given $f_{i}$, is more difficult. Instead of doing that, we used interpolation to give $B_{z,i}^{\text{exp}}$ at any given $f_{i}$. In other words, we first set a $B_{z,i}^{\text{cal}}$ to calculate a hyperfine transition frequency $f_i$, and then we want to compare this $B_{z,i}^{\text{cal}}$ with a $B_{z,i}^{\text{exp}}$. But since there is no such a $B_{z,i}^{\text{exp}}$ at that particular $f_i$, an interpolation from the closest experimental points is used to find a `synthetic' experimental point $B_{z,i}^{\text{exp}}$, and then the misfit at $f_{i}$ was calculated using Eq.~(\ref{eqs:misfit}). This method works for our case because the error of interpolation is much less than the uncertainties of our experiment. Using frequency instead of magnetic field to define a misfit function similar to Eq.~(\ref{eqs:misfit}) could be an alternative, but this requires carefully choosing different weighted numbers for data sets of different effective $g$ factors (approximately from 2 to 15 in our experiment). This is because different $g$ factors can introduce significantly different shifts in frequency even for the same error in magnetic field.

Using trial spin Hamiltonian parameters, we calculated points $(f_{i}, \, B_{z,i}^{\text{cal}})$ by diagonalization of the spin Hamiltonian. For each calculated magnetic field, the diagonalization of the spin Hamiltonian results in 120 possible transitions, and only those strong transitions could be seen in our EPR experiments, therefore those $f_i$ with a transition strength above one seventh of their maxima were considered to be measurable in our measurements (EPR signals are proportional to the squares of the transition strengths). The screening of weak transitions made the fitting process easier and faster. By using the least-squares method to minimize the misfit function in Eq.~(\ref{eqs:misfit}), our spin Hamiltonian parameters, together with $\theta$s, were calculated. The obtained spin Hamiltonian parameters were then refined by another run of weighted least-squares which took into account the zero field EPR data in literature \cite{Chen2016}. The results are given in Table~\ref{tab:vals}. 

After the best fit parameters were obtained, the ``temperature'' of the simulated annealing algorithm was raised to a level set by assuming each $B_{z,i}^{\text{exp}}$ has an uncertainty of 0.5\,mT. The result is a set of parameters that sample the posterior probability distribution. Uncertainties shown in Table~\ref{tab:vals} were calculated from the accepted 37652 samples. It is worth of noting that using the Markov-Chain-Monte-Carlo technique to find the uncertainties is difficult in our case due to the in total 38 parameters in the fitting; therefore the uncertainties for those numbers with big uncertainties might be underestimated.  For calculation purposes, we keep two significant digits for all the uncertainties and the fitted numbers are rounded accordingly. The experimental $\theta$ parameters given by fitting are  $\theta_1 = 64.04^\circ \pm 0.85^\circ$, $\theta_2 = 7.10^\circ \pm 0.77^\circ$, $\theta_3 = 104.04 ^\circ \pm 0.50^\circ$, $\theta_4 = 84.96 ^\circ \pm 0.41^\circ$. The resulting values for $\theta$ were in a range of values consistent with the photographic measurements. 

The spin Hamiltonian parameter matrices of site 1 and site 2 in $(\mathbf{D_1}, \mathbf{D_2}, \mathbf{b})$ obtained were: \\

\begin{equation}
\mathbf{g}_1 = \begin{bmatrix} 2.90 & -2.95 & -3.56 \\ -2.95 & 8.90 & 5.57 \\ -3.56 & 5.57 & 5.12 \end{bmatrix}
\end{equation} 
\begin{equation}
\mathbf{g}_2 = \begin{bmatrix} 14.37 & -1.77 & 2.40 \\ -1.77 & 1.93 & -0.43 \\ 2.40 & -0.43 & 1.44 \end{bmatrix}
\end{equation}
\begin{equation}
\mathbf{A}_1 = \begin{bmatrix} 274.3 & -202.5 & -350.8 \\ -202.5 & 827.5 & 635.2 \\ -350.8 & 635.2 & 706.1 \end{bmatrix} \text{MHz}
\end{equation} 
\begin{equation}
\mathbf{A}_2 = \begin{bmatrix} -1565.3 & 219.0 & -124.4 \\ 219.0 & -15.3 & -0.7 \\ -124.4 & -0.7 & 127.8 \end{bmatrix} \text{MHz}
\end{equation}
and the quadrupole interaction parameters are:
\begin{equation}
\mathbf{Q}_1 = \begin{bmatrix} 10.4 & -9.1 & -10.0 \\ -9.1 & -6.0 & -14.3 \\ -10.0 & -14.3 & -4.4 \end{bmatrix} \text{MHz}
\end{equation}
\begin{equation}
\mathbf{Q}_2 = \begin{bmatrix} -10.5 & -22.8 & -3.1 \\ -22.8 & -19.5 & -17.7 \\ -3.1 & -17.7 & 30.0 \end{bmatrix} \text{MHz}
\end{equation}

\begin {table*}[]
\caption{Principal values and Euler angles of $\textbf{g}$, $\textbf{A}$, and $\textbf{Q}$ of \er in \YSO \ at 4.3K.\label{tab:vals}}\bigskip
\begin{tabular}{l@{}|c@{}c@{}c|| l@{}|c@{}c@{}c}
 \hline \hline
   \multicolumn{4}{c||}{\multirow{2}{*}{Site 1}} & \multicolumn{4}{c}{\multirow{2}{*}{Site 2}} \\ 
    \multicolumn{4}{c||}{\multirow{2}{*}{}} & \multicolumn{4}{c}{\multirow{2}{*}{}}
      \\ 
 \hline  
 \multirow{3}{*}{Principal values} & & & & \multirow{3}{*}{ Principal values}& && \\
 & \multicolumn{3}{c||}{Euler angles (deg)} &  &\multicolumn{3}{c}{Euler angles (deg)} \\ 
 & \multicolumn{1}{c}{$\alpha$} & \multicolumn{1}{c}{$\beta$} & \multicolumn{1}{c||}{$\gamma$} & & \multicolumn{1}{c}{$\alpha$} & \multicolumn{1}{c}{$\beta$} & \multicolumn{1}{c}{$\gamma$} \\ 
\hline                                 
 $\begin{array}{@{}l} \\g_x = 2.253 \pm 0.049\\  g_y  = 0.057 \pm 0.26  \\  g_z = 14.61 \pm 0.10 \\ \\ \end{array}$  & $\begin{array}{c@{}}205.87 \\ \pm 0.43 \end{array}$ & $\begin{array}{c@{}} 55.33 \\ \pm 0.41 \end{array}$ & $ \begin{array}{c@{}} 29.3 \\ \pm 0.74 \end{array} $ & $\begin{array}{@{}l} \,g_x  = 0.99 \pm 0.37 \\  \,g_y  = 1.695 \pm 0.040 \\  \,g_z = 15.05 \pm 0.20 \end{array}$ & $ \begin{array}{c@{}} 261.97 \\ \pm 0.22 \end{array}$ & $\begin{array}{c@{}} 100.16 \\ \pm 1.1 \end{array}$ & $\begin{array}{c@{}} 97.25 \\ \pm 2.3 \end{array}$ \\ 
 \hline   
 $\begin{array}{@{}l} \\A_x = 260.1 \pm 11 \\  A_y  = 22 \pm 45 \quad \text{(MHz)} \\  A_z = 1524.9 \pm 4 \\  \\ \end{array}$ & $ \begin{array}{c@{}} 203.3 \\ \pm 1.2 \end{array} $ & $ \begin{array}{c@{}} 48.59 \\ \pm 1.14 \end{array} $ & $ \begin{array}{c@{}} 26.2 \\ \pm 2.4 \end{array}$ & $\begin{array}{@{}l} \\ \,A_x =  139 \pm 82\\ \, A_y  = 13 \pm 98  \quad \text{(MHz)} \\  \,A_z = -1604 \pm 20 \\ \\  \end{array}{}$ & $ \begin{array}{c@{}} 262.2 \\ \pm 0.98 \end{array}{} $ & $ \begin{array}{c@{}} 94.07 \\ \pm 6.0 \end{array}{} $ & $ \begin{array}{c@{}}81.9 \\ \pm 15.9 \end{array}{} $   \\ 
 \hline  
$\begin{array}{@{}l@{}@{}l} \\Q_x =  9.03 \pm 0.95\\  Q_y  = 15.7 \pm 1.5 & \text{(MHz)} \\  Q_z = -(Q_x+Q_y) \\ \\  \end{array}{}$ & $ \begin{array}{c@{}} 151.87 \\ \pm 3.0 \end{array}{} $  & $ \begin{array}{c@{}} 49.48 \\ \pm 5.2 \end{array}{} $ & $ \begin{array}{c@{}} 296.67 \\ \pm 8.9 \end{array}{} $ & $\begin{array}{@{}l@{}@{}l} \\ \,Q_x = 6 \pm 14 \\  \,Q_y  = 36.1 \pm 7.1 & \text{(MHz)}\\  \, Q_z = -(Q_x+Q_y) \\ \\  \end{array}{}$ & $ \begin{array}{c@{}} 142.81 \\ \pm 7.3 \end{array}{} $ & $ \begin{array}{c@{}} 77.5 \\ \pm 4.6 \end{array}{} $ & $ \begin{array}{c@{}} 16.8 \\ \pm 6.5 \end{array}{} $   \\ 
 \hline \hline
 \end{tabular} 
\end{table*}

Our magnets were calibrated with an accuracy of $\pm$0.5\% and  due to the precision of the current supplies that drive our superconducting magnets, we expect another uncertainty of $\pm$0.1\,mT on the measured magnetic field strength. Besides this we estimate the cables connected to the microwave resonator, which have magnetic components nearby, also distorted the magnetic field by approximately $\pm$0.3\%. The total Zeeman energy shift is then calculated to be $\mathcal{O}\text{(20\,MHz)}$, which agrees with the uncertainties of the principal values of $\textbf{A}$ and $\textbf{Q}$ in Table~\ref{tab:vals}. Another source of uncertainty is the errors in $\theta$. The initial values of $\theta$ were measured by the use of a camera, which has an accuracy of approximately $3^\circ$. These numbers were allowed to vary during the fitting processes, which yielded uncertainties of $\mathcal{O}(1^\circ)$. 

The ground state $g$ factors have been measured previously\cite{Guillot-Noel2006}, \cite{Sun2008}. The values of the $g$ factors determined from our experiment are similar to their measurements. As listed in Table~\ref{tab:vals}, the $g_z$ values are the most accurate which have uncertainties less than 2\%, and the $g_x$ values could not be well determined because the values themselves are small. Similarly, the largest principal values $A_z$ of both site 1 and site 2 are then the most accurate among the three principal values of the hyperfine interaction $\textbf{A}$, which have uncertainties of less than 1\%. The uncertainties of the principal values of $\textbf{A}$, $\mathcal{O}\text{(20\,MHz)}$, are limited by the error of our experiment, which makes the small principal values, $A_y$ less accurate, as shown in Table~\ref{tab:vals}. As for the quadrupole interaction $\textbf{Q}$, their uncertainties are comparable to their principal values, as listed in Table~\ref{tab:vals}. As a result, $\textbf{Q}$ was not as well determined as $\textbf{g}$ and $\textbf{A}$ in our experiment.

\begin{figure}
\includegraphics[width=0.45\textwidth]{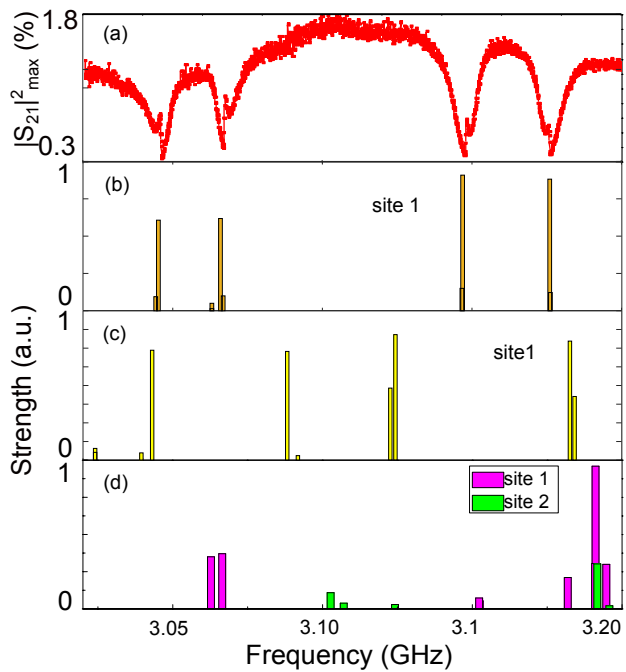}
\caption{(a) Experimental zero-field EPR spectra reproduced from the reference \cite{Chen2016}; (b) Calculated zero-field EPR spectrum with spin Hamiltonian parameters in Table~\ref{tab:vals}, which were obtained by fitting both the 2D EPR data and the zero-field EPR data. (c) Calculated zero-field EPR spectrum with spin Hamiltonian parameters which are obtained by just fitting the 2D EPR data. The transition strengths of site 2 in (b) and (c) are much smaller compared to site 1; therefore for clarity the transitions from site 2 are not shown in (b) and (c). (d) Calculated zero-field EPR spectrum with spin Hamiltonian parameters in the reference\cite{Guillot-Noel2006} . }
\label{fig:compZF}
\end{figure}

\begin{figure}
\includegraphics[width=0.45\textwidth]{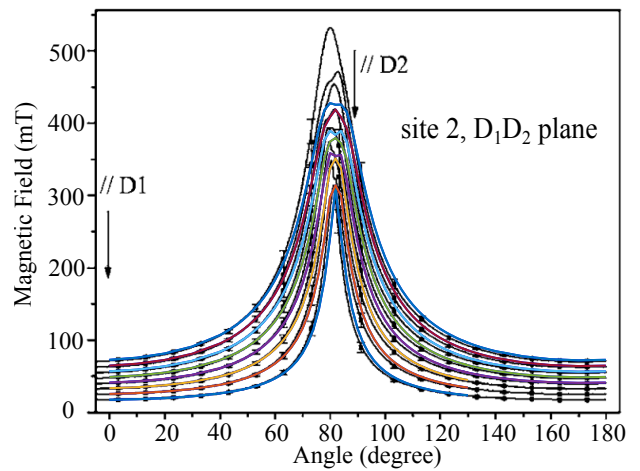}
\caption{Angular variation in the ($\mathbf{D_1,D_2}$) plane of eight allowed hyperfine transitions of site 2 at 9.5 GHz. Black spots and black lines are experimental points and fit curves adapted from the reference \cite{Guillot-Noel2006}; the coloured lines are transition lines based on spin Hamiltonian parameters in Table~\ref{tab:vals}. Similar behavior was observed when comparing our spin Hamiltonian with other results from the paper, except for one field rotation in the ($\mathbf{D_2,b'}$) plane where we were unsure of the exact orientation used.
}
\label{fig:compprb}
\end{figure}

\section{Reliability of Spin Hamiltonian Parameters}

When deriving the physical parameters of a system from fits to data, a potential problem is that the resulting fit is fortuitous. The fact that the derived parameters fit the data doesn't mean the ``true'' set of parameters have been found. This problem is exacerbated in a situation like \ersample\ because the low symmetry means that there is a large number of parameters. Furthermore the large nuclear spin of $^{167}$Er means that the spectra are in general very dense with spectral lines, and high precision is required if the spin Hamiltonian is going to be helpful in assigning lines.

As a test of our new spin Hamiltonian we made a prediction of the zero-field transitions reported previously\cite{Chen2016} (Fig.~\ref{fig:compZF}). Before we added this data to the fit the predicted spectra was consistent with the observed spectra given the $\sim$20\,MHz uncertainty in the spin Hamiltonian parameters. When the zero-field transitions were added to the dataset used for fitting, very good agreement was achieved. This indicates that while our spin Hamiltonian parameters can be used to successfully describe low magnetic field behaviour more data near zero-field could significantly improve the predictive power of the spin Hamiltonian parameters. In FIG. 4, the angular variation in the ($\mathbf{D_1,D_2}$) plane of eight allowed hyperfine transitions of site 2 predicted by the parameters in Table \ref{tab:vals} are plot on top of the experimental data from the literature\cite{Guillot-Noel2006}. Although the parameters in Table \ref{tab:vals} were obtained in $3 \sim 5$\,GHz, good agreements can be seen at higher frequency of 9.5 GHz. A more detailed comparison can be found in the Supplemental Material\cite{SupplMat}.

This leads to the question of why it is difficult to determine usable spin Hamiltonian parameters in this case. We believe it is related to the fact that the $g$ tensors in this case are very anisotropic, as shown in Table \ref{tab:vals}. When making standard high-field EPR measurements as was the case in the literature\cite{Guillot-Noel2006}, the dominant term in the spin Hamiltonian is $\mu_e \mathbf{B} \cdot \mathbf{g} \cdot \mathbf{S}$. This term splits the energy levels into two subspaces each of eight levels. One of theses subspaces has electron  spin up and the other has spin down along a quantization direction determined by the ``effective'' magnetic field direction  $\mathbf{\hat n}= \mathbf{B} \cdot \mathbf{g}/(|\mathbf{B} \cdot \mathbf{g}|)$. In each of these two subspaces $\mathbf{S}$ can be replaced with a classical vector pointing along or opposite to $\mathbf{\hat{n}}$. Because of this the next most significant term, which is the hyperfine splitting $\mathbf{I} \cdot \mathbf{A} \cdot \mathbf{S}$, is analogous to an anisotropic nuclear Zeeman term with $\mathbf{S}$ being like the applied magnetic field and $\mathbf{A}$ being like the Zeeman tensor. The nuclear Zeeman component of the energy eigenstates are therefore quantized along the direction $\mathbf{\hat{m}}=\mathbf{A}\cdot\mathbf{\hat{n}}/(|\mathbf{A}\cdot\mathbf{\hat{n}}|)$. Their splittings are determined primarily by $|\mathbf{A}\cdot\mathbf{\hat{n}} |$ with additional perturbations from the (real) nuclear Zeeman term and the nuclear quadrupole term.
In order to properly determine the hyperfine tensor $\mathbf{A}$, sufficient data must be collected such that the effective magnetic field $(\mathbf{\hat{n}})$ sufficiently samples all directions. This is made difficult by the fact that the $\mathbf{g}$ tensor is highly anisotropic. For most magnetic field directions, $\mathbf{\hat{n}}$ will end up close to pointing along the principle axis of the $\mathbf{g}$ tensor with the biggest principle value. This is particularly an issue for site 1 where one of the transverse $g$ values is very close to zero. A similar argument says that a good coverage of $\mathbf{\hat{m}}$ is needed to properly determine the $\mathbf{Q}$ tensor.

This problem with sufficient coverage of magnetic field directions is illustrated in the results of Table~\ref{tab:vals}. For hyperfine $\mathbf{A}$, the biggest remaining uncertainty for site 1 is $A_y$ that has its principle direction close to the principle axis of $g_y$ whose value is close to zero. It can also be seen manifested in Fig.~\ref{fig:compprb} where predictions from our new spin Hamiltonian parameters are overlaid on a figure from the reference \cite{Guillot-Noel2006}. The experimental data agrees with both the new and the old spin Hamiltonian predictions. However the old and new predictions differ significantly for the largest magnetic field values, which is apparent because the large magnetic field values correspond to a small $g$ value at fixed-frequency EPR experiments.

For this work a tunable resonator was chosen to overcome these difficulties, by allowing measurements where energy levels are anti-crossing near zero magnetic field. This wasn't entirely successful because the signals for both standard EPR and the variant used here disappear in these situations.


Almost all the data used here were straight lines on the magnetic field versus frequency graphs (see Fig.~\ref{fig:spectra}). The improvement of our parameters over others can be mostly attributed as much to better coverage of effective magnetic field directions as the fact that energy levels are highly mixed at low magnetic field.

\section{Degeneracy of spin Hamiltonian Parameters}

Another issue is that many different spin Hamiltonians lead to the same EPR behaviour. Given a vector of spin operators $\mathbf{S}=(S_x,S_y,S_z)$ applying a rotation $\mathbf{U}$ to this vector of operators gives a set of spin operators with the same algebra. For this reason given an arbitrary rotation  $\mathbf{U}$, Eq.~(\ref{eq:spinHam}) and \begin{equation}
H=\mu_e \mathbf{B} \cdot \mathbf{g} \cdot \mathbf{U} \cdot \mathbf{S} + \mathbf{I} \cdot \mathbf{A} \cdot \mathbf{U} \cdot \mathbf{S} +\mathbf{I}  \cdot \mathbf{Q}  \cdot \mathbf{I} -\mu_n g_n \mathbf{B} \cdot\mathbf{I} ,
\end{equation} 
are equivalent. 

Because the nuclear Zeeman term is very small given another arbitrary rotation $\mathbf{V}$ leads to another set of almost equivalent Hamiltonians\begin{equation}
H=\mu_e \mathbf{B} \cdot \mathbf{g} \cdot \mathbf{U} \cdot \mathbf{S} + \mathbf{
I} \cdot \mathbf{V}^\text{T} \cdot \mathbf{A} \cdot \mathbf{U} \cdot \mathbf{S} 
+\mathbf{I} \cdot \mathbf{V}^\text{T} \cdot \mathbf{Q}  \cdot \mathbf{V} \cdot\mathbf{I} -\mu_n g_n \mathbf{B} \cdot \mathbf{V} \cdot\mathbf{I} ,
\end{equation}

This degeneracy does mean that one has to be careful before saying that two spin spin Hamiltonians are really different, and it is important if one is trying to understand the nature of the site at a deeper level. As the predictions for energy levels and EPR transition strengths under these rotations are the same, it wasn't the reason for the discrepancies addressed in this work. Furthermore the standard practice of taking the matrices to be symmetric, means that this degeneracy only effects the sign of the principle values. 

The spin Hamiltonian Eq.~(\ref{eq:spinHam}) won't be valid for arbitarily large magnetic field values, because at high fields the upper electron spin manifold from the ground crystal-field level (Z$_1$) will start to mix with the lower electron spin manifold from the second lowest crystal-field levels (Z$_2$). We don't think that this is the reason for the discrepancy between this work and the old spin parameters for two reasons. Firstly, Z$_2$ is 1260\,GHz higher in energy than Z$_1$ \cite{DouaIan1995} and the measurements in the reference \cite{Guillot-Noel2006} were made with only 9.5 GHz splittings so the mixing between the levels should be small. Our crystal field calculations support this assertion. The second reason is that our new spin hamiltonian agrees with the measurements from the reference \cite{Guillot-Noel2006}. It only disagrees with the spin Hamiltonian parameters that was derived from these measurements.

\section{Conclusion}
In conclusion, we have characterized the hyperfine structure of the ground state of \er \ ions in \YSO \ by measuring EPR spectra in a tunable microwave cavity for different crystal orientations. Compared to conventional EPR, the ability to vary the resonator frequency and magnetic field gives more details of how the frequency of one particular hyperfine transition depends on the applied magnetic field in a 2D pattern. Based on the 2D EPR data, the matrices of the Zeeman $g$ factor, hyperfine interactions, and quadrupole interactions of \ersample \ are determined by least-squares fitting.  The uncertainties of predicted hyperfine energy levels are approximately 20\,MHz. The spin Hamiltonian parameters not only agree with the 2D EPR scans in this paper, but are also consistent with the zero-field EPR data \cite{Chen2016} and the EPR data at 9.5\,GHz\cite{Guillot-Noel2006}. The difficulty in characterizing the hyperfine structure of \ersample \ is ascribed to the fact that $\mathbf{g}$ tensors are highly anisotropic which means special attention must be given to the  coverage of magnetic field directions. To further narrow down the spin Hamiltonian parameters, transition points at zero magnetic field or anti-crossing points at low magnetic fields would be useful. While standard  microwave EPR spectroscopy is no longer suitable for anti-crossing points, the combination of both optical and microwave detection, e.g. Raman Heterodyne spectroscopy\cite{Fernandez-Gonzalvo2015a}, is a possibility.

\section{Acknowledgements}
The authors would like to thank Mike Reid and Philippe Goldner for useful discussions. This work was supported by the Marsden Fund of the Royal Society of New Zealand through contract UoO1520.


\begin{thebibliography}{21}
\expandafter\ifx\csname natexlab\endcsname\relax\def\natexlab#1{#1}\fi
\expandafter\ifx\csname bibnamefont\endcsname\relax
  \def\bibnamefont#1{#1}\fi
\expandafter\ifx\csname bibfnamefont\endcsname\relax
  \def\bibfnamefont#1{#1}\fi
\expandafter\ifx\csname citenamefont\endcsname\relax
  \def\citenamefont#1{#1}\fi
\expandafter\ifx\csname url\endcsname\relax
  \def\url#1{\texttt{#1}}\fi
\expandafter\ifx\csname urlprefix\endcsname\relax\def\urlprefix{URL }\fi
\providecommand{\bibinfo}[2]{#2}
\providecommand{\eprint}[2][]{\url{#2}}

\bibitem[{\citenamefont{Thiel et~al.}(2011)\citenamefont{Thiel, B{\"{o}}ttger,
  and Cone}}]{Thiel2011}
\bibinfo{author}{\bibfnamefont{C.}~\bibnamefont{Thiel}},
  \bibinfo{author}{\bibfnamefont{T.}~\bibnamefont{B{\"{o}}ttger}},
  \bibnamefont{and} \bibinfo{author}{\bibfnamefont{R.}~\bibnamefont{Cone}},
  \bibinfo{journal}{J. Lumin.} \textbf{\bibinfo{volume}{131}},
  \bibinfo{pages}{353} (\bibinfo{year}{2011}), ISSN \bibinfo{issn}{00222313},
  \urlprefix\url{http://linkinghub.elsevier.com/retrieve/pii/S002223131000534X}.

\bibitem[{\citenamefont{Zhong et~al.}(2015)\citenamefont{Zhong, Hedges,
  Ahlefeldt, Bartholomew, Beavan, Wittig, Longdell, and Sellars}}]{Zhong2015}
\bibinfo{author}{\bibfnamefont{M.}~\bibnamefont{Zhong}},
  \bibinfo{author}{\bibfnamefont{M.~P.} \bibnamefont{Hedges}},
  \bibinfo{author}{\bibfnamefont{R.~L.} \bibnamefont{Ahlefeldt}},
  \bibinfo{author}{\bibfnamefont{J.~G.} \bibnamefont{Bartholomew}},
  \bibinfo{author}{\bibfnamefont{S.~E.} \bibnamefont{Beavan}},
  \bibinfo{author}{\bibfnamefont{S.~M.} \bibnamefont{Wittig}},
  \bibinfo{author}{\bibfnamefont{J.~J.} \bibnamefont{Longdell}},
  \bibnamefont{and} \bibinfo{author}{\bibfnamefont{M.~J.}
  \bibnamefont{Sellars}}, \bibinfo{journal}{Nature}
  \textbf{\bibinfo{volume}{517}}, \bibinfo{pages}{177} (\bibinfo{year}{2015}),
  ISSN \bibinfo{issn}{0028-0836}, \eprint{1411.6758},
  \urlprefix\url{http://www.nature.com/nature/journal/v517/n7533/full/nature14025.html{\%}5Cnhttp://www.nature.com/nature/journal/v517/n7533/pdf/nature14025.pdf}.

\bibitem[{\citenamefont{Fraval et~al.}(2004)\citenamefont{Fraval, Sellars, and
  Longdell}}]{Fraval2005}
\bibinfo{author}{\bibfnamefont{E.}~\bibnamefont{Fraval}},
  \bibinfo{author}{\bibfnamefont{M.~J.} \bibnamefont{Sellars}},
  \bibnamefont{and} \bibinfo{author}{\bibfnamefont{J.~J.}
  \bibnamefont{Longdell}}, \bibinfo{journal}{Phys. Rev. Lett.}
  \textbf{\bibinfo{volume}{95}}, \bibinfo{pages}{030506}
  (\bibinfo{year}{2005}), 
  \urlprefix\url{https://journals.aps.org/prl/abstract/10.1103/PhysRevLett.95.030506}.

\bibitem[{\citenamefont{Macfarlane et~al.}(1997)\citenamefont{Macfarlane,
  Harris, Sun, Cone, and Equall}}]{Macfarlane1997}
\bibinfo{author}{\bibfnamefont{R.~M.} \bibnamefont{Macfarlane}},
  \bibinfo{author}{\bibfnamefont{T.~L.} \bibnamefont{Harris}},
  \bibinfo{author}{\bibfnamefont{Y.}~\bibnamefont{Sun}},
  \bibinfo{author}{\bibfnamefont{R.~L.} \bibnamefont{Cone}}, \bibnamefont{and}
  \bibinfo{author}{\bibfnamefont{R.~W.} \bibnamefont{Equall}},
  \bibinfo{journal}{Opt. Lett.} \textbf{\bibinfo{volume}{22}},
  \bibinfo{pages}{871} (\bibinfo{year}{1997}), ISSN \bibinfo{issn}{0146-9592}.

\bibitem[{\citenamefont{B{\"{o}}ttger et~al.}(2006)\citenamefont{B{\"{o}}ttger,
  Thiel, Sun, and Cone}}]{Bottger2006}
\bibinfo{author}{\bibfnamefont{T.}~\bibnamefont{B{\"{o}}ttger}},
  \bibinfo{author}{\bibfnamefont{C.~W.}~\bibnamefont{Thiel}},
  \bibinfo{author}{\bibfnamefont{Y.}~\bibnamefont{Sun}}, \bibnamefont{and}
  \bibinfo{author}{\bibfnamefont{R.~L.}~\bibnamefont{Cone}},
  \bibinfo{journal}{Phys. Rev. B} \textbf{\bibinfo{volume}{73}},
  \bibinfo{pages}{075101} (\bibinfo{year}{2006}), ISSN
  \bibinfo{issn}{1098-0121},
  \urlprefix\url{http://link.aps.org/doi/10.1103/PhysRevB.73.075101}.

\bibitem[{\citenamefont{McAuslan et~al.}(2012)\citenamefont{McAuslan,
  Bartholomew, Sellars, and Longdell}}]{McAuslan2012}
\bibinfo{author}{\bibfnamefont{D.~L.} \bibnamefont{McAuslan}},
  \bibinfo{author}{\bibfnamefont{J.~G.} \bibnamefont{Bartholomew}},
  \bibinfo{author}{\bibfnamefont{M.~J.} \bibnamefont{Sellars}},
  \bibnamefont{and} \bibinfo{author}{\bibfnamefont{J.~J.}
  \bibnamefont{Longdell}}, \bibinfo{journal}{Phys. Rev. A - At. Mol. Opt.
  Phys.} \textbf{\bibinfo{volume}{85}} (\bibinfo{year}{2012}), ISSN
  \bibinfo{issn}{10502947}, \eprint{arXiv:1201.4610v1}.

\bibitem[{\citenamefont{Williamson et~al.}(2014)\citenamefont{Williamson, Chen,
  and Longdell}}]{Williamson2014a}
\bibinfo{author}{\bibfnamefont{L.~A.} \bibnamefont{Williamson}},
  \bibinfo{author}{\bibfnamefont{Y.-H.} \bibnamefont{Chen}}, \bibnamefont{and}
  \bibinfo{author}{\bibfnamefont{J.~J.} \bibnamefont{Longdell}},
  \bibinfo{journal}{Phys. Rev. Lett.} \textbf{\bibinfo{volume}{113}},
  \bibinfo{pages}{203601} (\bibinfo{year}{2014}), ISSN
  \bibinfo{issn}{0031-9007},
  \urlprefix\url{http://link.aps.org/doi/10.1103/PhysRevLett.113.203601}.

\bibitem[{\citenamefont{Chen et~al.}(2016)\citenamefont{Chen,
  Fernandez-Gonzalvo, and Longdell}}]{Chen2016}
\bibinfo{author}{\bibfnamefont{Y.~H.} \bibnamefont{Chen}},
  \bibinfo{author}{\bibfnamefont{X.}~\bibnamefont{Fernandez-Gonzalvo}},
  \bibnamefont{and} \bibinfo{author}{\bibfnamefont{J.~J.}
  \bibnamefont{Longdell}}, \bibinfo{journal}{Phys. Rev. B - Condens. Matter
  Mater. Phys.} \textbf{\bibinfo{volume}{94}}, \bibinfo{pages}{075117}
  (\bibinfo{year}{2016}), ISSN \bibinfo{issn}{1550235X}, \eprint{1512.03606}.

\bibitem[{\citenamefont{Baldit et~al.}(2010)\citenamefont{Baldit, Bencheikh,
  Monnier, Briaudeau, Levenson, Crozatier, Lorger{\'{e}}, Bretenaker, {Le
  Gou{\"{e}}t}, Guillot-No{\"{e}}l et~al.}}]{Baldit2010}
\bibinfo{author}{\bibfnamefont{E.}~\bibnamefont{Baldit}},
  \bibinfo{author}{\bibfnamefont{K.}~\bibnamefont{Bencheikh}},
  \bibinfo{author}{\bibfnamefont{P.}~\bibnamefont{Monnier}},
  \bibinfo{author}{\bibfnamefont{S.}~\bibnamefont{Briaudeau}},
  \bibinfo{author}{\bibfnamefont{J.~a.} \bibnamefont{Levenson}},
  \bibinfo{author}{\bibfnamefont{V.}~\bibnamefont{Crozatier}},
  \bibinfo{author}{\bibfnamefont{I.}~\bibnamefont{Lorger{\'{e}}}},
  \bibinfo{author}{\bibfnamefont{F.}~\bibnamefont{Bretenaker}},
  \bibinfo{author}{\bibfnamefont{J.~L.} \bibnamefont{{Le Gou{\"{e}}t}}},
  \bibinfo{author}{\bibfnamefont{O.}~\bibnamefont{Guillot-No{\"{e}}l}},
  \bibnamefont{et~al.}, \bibinfo{journal}{Phys. Rev. B}
  \textbf{\bibinfo{volume}{81}}, \bibinfo{pages}{144303}
  (\bibinfo{year}{2010}), ISSN \bibinfo{issn}{1098-0121},
  \urlprefix\url{http://link.aps.org/doi/10.1103/PhysRevB.81.144303}.

\bibitem[{\citenamefont{Hashimoto and Shimizu}(2016)}]{Hashimoto2016}
\bibinfo{author}{\bibfnamefont{D.}~\bibnamefont{Hashimoto}} \bibnamefont{and}
  \bibinfo{author}{\bibfnamefont{K.}~\bibnamefont{Shimizu}},
  \bibinfo{journal}{J. Lumin.} \textbf{\bibinfo{volume}{171}},
  \bibinfo{pages}{183} (\bibinfo{year}{2016}), ISSN \bibinfo{issn}{00222313},
  \urlprefix\url{http://dx.doi.org/10.1016/j.jlumin.2015.11.031}.

\bibitem[{\citenamefont{Ran{\v{c}}i{\'{c}}
  et~al.}(2016)\citenamefont{Ran{\v{c}}i{\'{c}}, Hedges, Ahlefeldt, and
  Sellars}}]{Rancic2016}
\bibinfo{author}{\bibfnamefont{M.}~\bibnamefont{Ran{\v{c}}i{\'{c}}}},
  \bibinfo{author}{\bibfnamefont{M.~P.} \bibnamefont{Hedges}},
  \bibinfo{author}{\bibfnamefont{R.~L.} \bibnamefont{Ahlefeldt}},
  \bibnamefont{and} \bibinfo{author}{\bibfnamefont{M.~J.}
  \bibnamefont{Sellars}}, \bibinfo{journal}{arXiv}  (\bibinfo{year}{2016}),
  \eprint{1611.04315}, \urlprefix\url{http://arxiv.org/abs/1611.04315}.

\bibitem[{\citenamefont{Asatryan and Rosa}(2002)}]{Asatryan2002}
\bibinfo{author}{\bibfnamefont{G.~R.} \bibnamefont{Asatryan}} \bibnamefont{and}
  \bibinfo{author}{\bibfnamefont{J.}~\bibnamefont{Rosa}},
  \bibinfo{journal}{Phys. Solid State} \textbf{\bibinfo{volume}{44}},
  \bibinfo{pages}{864} (\bibinfo{year}{2002}).

\bibitem[{\citenamefont{Jerzak and Mroz}(2003)}]{Jerzak2003}
\bibinfo{author}{\bibfnamefont{S.}~\bibnamefont{Jerzak}} \bibnamefont{and}
  \bibinfo{author}{\bibfnamefont{B.}~\bibnamefont{Mroz}}, \bibinfo{journal}{J.
  Phys. Condens. MATTER} \textbf{\bibinfo{volume}{15}}, \bibinfo{pages}{5113}
  (\bibinfo{year}{2003}).

\bibitem[{\citenamefont{Guillot-No{\"{e}}l
  et~al.}(2006)\citenamefont{Guillot-No{\"{e}}l, Goldner, Du, Baldit, Monnier,
  and Bencheikh}}]{Guillot-Noel2006}
\bibinfo{author}{\bibfnamefont{O.}~\bibnamefont{Guillot-No{\"{e}}l}},
  \bibinfo{author}{\bibfnamefont{P.}~\bibnamefont{Goldner}},
  \bibinfo{author}{\bibfnamefont{Y.~L.}~\bibnamefont{Du}},
  \bibinfo{author}{\bibfnamefont{E.}~\bibnamefont{Baldit}},
  \bibinfo{author}{\bibfnamefont{P.}~\bibnamefont{Monnier}}, \bibnamefont{and}
  \bibinfo{author}{\bibfnamefont{K.}~\bibnamefont{Bencheikh}},
  \bibinfo{journal}{Phys. Rev. B} \textbf{\bibinfo{volume}{74}},
  \bibinfo{pages}{214409} (\bibinfo{year}{2006}), ISSN
  \bibinfo{issn}{1098-0121},
  \urlprefix\url{http://link.aps.org/doi/10.1103/PhysRevB.74.214409}.

\bibitem[{\citenamefont{Sun et~al.}(2008)\citenamefont{Sun, B{\"{o}}ttger,
  Thiel, and Cone}}]{Sun2008}
\bibinfo{author}{\bibfnamefont{Y.}~\bibnamefont{Sun}},
  \bibinfo{author}{\bibfnamefont{T.}~\bibnamefont{B{\"{o}}ttger}},
  \bibinfo{author}{\bibfnamefont{C.~W.}~\bibnamefont{Thiel}}, \bibnamefont{and}
  \bibinfo{author}{\bibfnamefont{R.~L.}~\bibnamefont{Cone}},
  \bibinfo{journal}{Phys. Rev. B} \textbf{\bibinfo{volume}{77}},
  \bibinfo{pages}{085124} (\bibinfo{year}{2008}), ISSN
  \bibinfo{issn}{1098-0121},
  \urlprefix\url{http://link.aps.org/doi/10.1103/PhysRevB.77.085124}.

\bibitem[{\citenamefont{Marino et~al.}(2016)\citenamefont{Marino,
  Lorger{\'{e}}, Guillot-No{\"{e}}l, Vezin, Toncelli, Tonelli, {Le
  Gou{\"{e}}t}, and Goldner}}]{Marino2016}
\bibinfo{author}{\bibfnamefont{R.}~\bibnamefont{Marino}},
  \bibinfo{author}{\bibfnamefont{I.}~\bibnamefont{Lorger{\'{e}}}},
  \bibinfo{author}{\bibfnamefont{O.}~\bibnamefont{Guillot-No{\"{e}}l}},
  \bibinfo{author}{\bibfnamefont{H.}~\bibnamefont{Vezin}},
  \bibinfo{author}{\bibfnamefont{A.}~\bibnamefont{Toncelli}},
  \bibinfo{author}{\bibfnamefont{M.}~\bibnamefont{Tonelli}},
  \bibinfo{author}{\bibfnamefont{J.-L.} \bibnamefont{{Le Gou{\"{e}}t}}},
  \bibnamefont{and} \bibinfo{author}{\bibfnamefont{P.}~\bibnamefont{Goldner}},
  \bibinfo{journal}{J. Lumin.} \textbf{\bibinfo{volume}{169}},
  \bibinfo{pages}{478} (\bibinfo{year}{2016}), ISSN \bibinfo{issn}{00222313},
  \urlprefix\url{http://linkinghub.elsevier.com/retrieve/pii/S0022231315001398}.

\bibitem[{\citenamefont{Aster, Borchers, and Thurber}(2011)}]{Aster2011}
\bibinfo{author}{\bibfnamefont{R.~C.}~\bibnamefont{Aster}},
\bibinfo{author}{\bibfnamefont{B.}~\bibnamefont{Borchers}}, \bibnamefont{and}  \bibinfo{author}{\bibfnamefont{C.~H.}~\bibnamefont{Therber}}, \bibinfo{book}{Parameter Estimation and Inverse Problems}, Academic Press (\bibinfo{year}{2011}).

\bibitem[{\citenamefont{{A. Abragam} and Pryce}(1951)}]{A.Abragam1951}
\bibinfo{author}{\bibnamefont{{A. Abragam}}} \bibnamefont{and}
  \bibinfo{author}{\bibfnamefont{M.~H.~L.} \bibnamefont{Pryce}},
  \bibinfo{journal}{Proc. R. Soc. A} \textbf{\bibinfo{volume}{205}},
  \bibinfo{pages}{135 } (\bibinfo{year}{1951}), ISSN \bibinfo{issn}{1364-5021}.

\bibitem[{\citenamefont{Fernandez-Gonzalvo
  et~al.}(2015)\citenamefont{Fernandez-Gonzalvo, Chen, Yin, Rogge, and
  Longdell}}]{Fernandez-Gonzalvo2015a}
\bibinfo{author}{\bibfnamefont{X.}~\bibnamefont{Fernandez-Gonzalvo}},
  \bibinfo{author}{\bibfnamefont{Y.~H.} \bibnamefont{Chen}},
  \bibinfo{author}{\bibfnamefont{C.}~\bibnamefont{Yin}},
  \bibinfo{author}{\bibfnamefont{S.}~\bibnamefont{Rogge}}, \bibnamefont{and}
  \bibinfo{author}{\bibfnamefont{J.~J.} \bibnamefont{Longdell}},
  \bibinfo{journal}{Phys. Rev. A - At. Mol. Opt. Phys.}
  \textbf{\bibinfo{volume}{92}}, \bibinfo{pages}{1} (\bibinfo{year}{2015}),
  ISSN \bibinfo{issn}{10941622}, \eprint{1501.02014}.

\bibitem[{\citenamefont{Goryachev et~al.}(2015)\citenamefont{Goryachev, Farr,
  Carvalho, Creedon, Floch, Probst, Bushev, and Tobar}}]{Goryachev2014a}
\bibinfo{author}{\bibfnamefont{M.}~\bibnamefont{Goryachev}},
  \bibinfo{author}{\bibfnamefont{W.~G.} \bibnamefont{Farr}},
  \bibinfo{author}{\bibfnamefont{N.~C.} \bibnamefont{Carvalho}},
  \bibinfo{author}{\bibfnamefont{D.~L.} \bibnamefont{Creedon}},
  \bibinfo{author}{\bibfnamefont{J.-M.~L.} \bibnamefont{Floch}},
  \bibinfo{author}{\bibfnamefont{S.}~\bibnamefont{Probst}},
  \bibinfo{author}{\bibfnamefont{P.}~\bibnamefont{Bushev}}, \bibnamefont{and}
  \bibinfo{author}{\bibfnamefont{M.~E.} \bibnamefont{Tobar}},
  \bibinfo{journal}{Appl. Phys. Lett.} \textbf{\bibinfo{volume}{106}},
  \bibinfo{pages}{232401} (\bibinfo{year}{2015}), \eprint{1410.6578},
  \urlprefix\url{http://arxiv.org/abs/1410.6578}.

\bibitem[{\citenamefont{Supplemental Material}(2017)}]{SupplMat}
{See Supplemental Material at [URL will be inserted by publisher] for a complete set of the data.}

\bibitem[{\citenamefont{DouaIan et~al.}(1995)\citenamefont{DouaIan, Labb\'{e}, Boulanger, Margerie, Moncorg\'{e} and Timonen}}]{DouaIan1995}
\bibinfo{author}{\bibfnamefont{J.L.}~\bibnamefont{DouaIan}},
  \bibinfo{author}{\bibfnamefont{C.} \bibnamefont{Labb\'{e}}},
  \bibinfo{author}{\bibfnamefont{P.~Le} \bibnamefont{Boulanger}},
  \bibinfo{author}{\bibfnamefont{J.} \bibnamefont{Margerie}},
  \bibinfo{author}{\bibfnamefont{R.} \bibnamefont{Moncorg\'{e}}}, \bibnamefont{and}
  \bibinfo{author}{\bibfnamefont{H.} \bibnamefont{Timonen}},
  \bibinfo{journal}{J. Phys.: Condens. Matter} \textbf{\bibinfo{volume}{7}},
  \bibinfo{pages}{5111} (\bibinfo{year}{1995}),
  \urlprefix\url{http://iopscience.iop.org/article/10.1088/0953-8984/7/26/017}.
  

\end{thebibliography}

\end{document}